 \documentclass[12pt]{iopart}

\usepackage[utf8]{inputenc}
\usepackage[T1]{fontenc}
\usepackage[english]{babel}

\usepackage{amsmath,amsfonts,amssymb,mathtools}
\usepackage {graphicx}
\graphicspath{{fig/}}
\usepackage{caption}

\usepackage{xcolor}

\usepackage{makecell}
\setcellgapes{4pt}
\usepackage{booktabs}
\usepackage{multirow}
\usepackage{graphicx}

\usepackage{cite}
\usepackage{hyperref}
\hypersetup{
	colorlinks=true,        
	linkcolor=blue,          
	citecolor=blue,        
	filecolor=magenta,      
}

\usepackage{array}
\usepackage{comment}

\allowdisplaybreaks[4]

\usepackage{color} 				
\definecolor{dgray}{rgb}{0.6,0.6,0.6}
\definecolor{dmag}{rgb}{0.6,0.0,0.6}
\definecolor{mbul}{rgb}{0.102, 0.42, 0.102} 
\usepackage[normalem]{ulem} 	

\definecolor{pink}{rgb}{1,0,0.9}

\newcommand{\pd}{{\phantom{\dagger}}}

\begin{document}

\title[]{\bf Zero sound in a quantum gas of spin-3/2 atoms with multipole exchange interaction}

\author{M Bulakhov$^{1,2,\dagger}$ , A S Peletminskii$^{1,2,*}$, 
and Yu~V~Slyusarenko$^{1,3}$}
\address{$^1$ Akhiezer Institute for Theoretical Physics, National Science Center "Kharkov Institute of Physics and Technology", NAS of Ukraine, 61108 Kharkiv, Ukraine}
\address{$^2$ V.N. Karazin Kharkiv National University, 61022 Kharkiv, Ukraine}
\address{$^3$ Lviv Polytechnic National University, 79000 Lviv, Ukraine}
\ead{$^\dagger$ \mailto{bulakh@kipt.kharkov.ua}}
\ead{$^*$ \mailto{aspelet@kipt.kharkov.ua}}



\begin{abstract}
In the context of quantum gases, we obtain a many-body Hamiltonian for spin-3/2 atoms with general multipole (spin, quadrupole, and octupole) exchange interaction by employing the apparatus of irreducible spherical tensor operators. This Hamiltonian implies the finite-range interaction, whereas, for zero-range (contact) potentials parameterized by the $s$-wave scattering length, the multipole exchange interaction becomes irrelevant. Following the reduced description method for quantum systems, we derive the quantum kinetic equation for spin-3/2 atoms in a magnetic field and apply it to examine the high-frequency oscillations known as zero sound. 
\end{abstract}

\maketitle

\section{Introduction}

Laser cooling and trapping techniques of neutral atoms \cite{Phillips1998,Tannoudji1998} elaborated a few decades ago have made a significant breakthrough in the physics of quantum gases. Due to them, one has succeeded in confirming the earlier theoretical predictions and shedding light on new physical effects and phenomena which are difficult to probe in natural materials \cite{PitStr2016,Pethick2008}. The uniqueness of confined ultracold dilute gases lies in a high degree of control of such physical parameters as interaction strength, density, temperature, spin, and even dimensionality. 

In contrast to purely magnetic trapping, at which the atomic spin degrees of freedom are frozen, the optical trapping gives the opportunity to examine all magnetic sublevels 
of spin-$F$ bosonic and fermionic atoms. In particular, as was demonstrated experimentally \cite{Stamper1998}, all three hyperfine states of spin-1 sodium atoms can be simultaneously Bose condensed in an optical trap. This experiment has stimulated a large number of theoretical \cite{Ohmi1998,HoPRL1998,JETP1998,Bigelow1998,UedaPRA2007} and experimental \cite{Ketterle1998,Stamper1999,Dalibard2012,Dalibard2019} studies of the so-called spinor Bose-Einstein condensates (for review articles, see \cite{UedaPhysRep,UedaRMP}). An amazing feature of such high-spin ($F>1/2$) quantum systems at ultralow temperature is that they can exhibit both magnetic and superfluid properties. 
As for high-spin Fermi gases, they attract much interest to reveal the effect of high atomic spin on the structure of Cooper pairs \cite{Ho1999}, spin waves \cite{Lewenstein2013}, zero sound \cite{HoYip1999}, etc.  
It is worth stressing that the interatomic interaction plays a crucial role in understanding and theoretical description of magnetic, superfluid and other collective phenomena in dilute quantum atomic gases. 

For low-energy atoms interacting through the short-range potential (those potentials, which decay algebraically at large distances with a power greater than the spatial dimension; e.g., the Van der Waals forces with decay law $1/r^{6}$ in 3D), one can replace it by a zero-range ($\delta$-like) pseudopotential, whose strength is specified by $s$-wave scattering length \cite{PitStr2016,Pethick2008}. Therefore, in spite of the fact that all realistic interactions are non-contact, the so-called “scattering-length approximation” neglects their finite range. This leads to a divergence of physical quantities, which is eliminated by renormalizing the coupling constant associated with $s$-wave scattering length \cite{PitStr2016,Pethick2008,Peletminskii_2017}. Nevertheless, the scattering-length approximation has proved to be a powerful tool while describing the short-range interaction effects in dilute quantum atomic gases \cite{PitStr2016,Pethick2008}. 
Within this approximation, the magnetic properties of spin-1 atomic gas with Bose-Einstein condensate have been studied by employing the Hamiltonian, which consists of two interaction terms \cite{Ohmi1998,HoPRL1998,UedaPhysRep}: the first term is independent of spin operators and the second term is bilinear in them (spin exchange interaction). However, if the finite range of interaction is essential in systems with total atomic spin $F>1/2$, then we are faced with the necessity to consider the multipole degrees of freedom. In particular, the quadrupole degrees of freedom should be included into the interaction Hamiltonian while studying the foregoing spin-1 condensate beyond the scattering-length approximation \cite{Peletminskii_2020,Bulakhov_2021,Bulakhov_2022}.
Therefore, the experimental observation of physical effects associated with multipole degrees of freedom must indicate the non-contact character of interatomic interaction in quantum gases. 
Moreover, since the quadrupole exchange is related to the long-range dipole-dipole interaction \cite{Racah1959}, for atoms with a sufficiently large dipole moment (like Erbium \cite{Aikawa_PRL2012,Ferlaino_PRA2022} and Dysprosium \cite{Lu_PRL2011,Lev_PRA2015}), the strengths of short-range and long-range interactions can be of the same order of magnitude \cite{PitStr2016,Bulakhov_2022}.

A natural question arises as to what collective phenomena in high-spin Fermi systems reveal multipole degrees of freedom. The high-frequency oscillations associated with the deviation of the single-particle density matrix from its equilibrium value can be considered for this role. These specific oscillations known as zero sound can propagate even at zero temperature.   
Initially, zero sound was theoretically predicted in liquid ${}^{3}$He \cite{Landau1957}. However, high-frequency oscillations at zero temperature can also propagate as spin waves in the electron liquid of metals \cite{Silin1_1958,Silin2_1958}. The studies were carried out within the Fermi-liquid theory, which implies the realization of the so-called Pomeranchuk stability conditions for the normal state \cite{Pomeranchuk_1958}. 
The zero-sound dispersion law can indicate their violation (the normal state becomes unstable), which signals a phase transition to another state \cite{LTP-II-1999,LTP-I-1999}. The experimental measuring of zero sound in liquid ${}^{3}$He \cite{Abel_1966,Roach_1976} has been one of the brightest confirmation of the celebrated Fermi-liquid theory.
As for ultracold Fermi gases, the zero sound modes were examined for high-spin atoms with contact interaction \cite{HoYip1999} as well as for atoms interacting through the long-range dipole-dipole forces \cite{Bohn2010,Shlyapnikov2013}.

In this paper, we address the ultracold interacting gas of spin-3/2 atoms in a magnetic field representing the most straightforward high-spin Fermi system and explore such a collective phenomenon as zero sound. To this end, we construct and justify a many-body Hamiltonian in terms of the irreducible spherical tensor operators, which allow us to include all  multipole (spin, quadrupole, and octupole) exchange interactions. This Hamiltonian is applied to obtain the relevant kinetic equation. Since we are interested in the dispersion law of high-frequency excitations close to the equilibrium state, we seek its wave solution in the collisionless approximation. 
In particular, we show that zero sound cannot propagate in a system with contact interaction. Therefore, its experimental observation in high-spin gases could serve as an indirect manifestation of multipole interactions inherent to systems with finite range forces.

\section{Pairwise interaction Hamiltonian of spin-3/2 atoms }

Consider the interaction of two identical spin-$F$ atoms. In general case, in such high-spin system, there is an exchange of any multipole moment. In order to construct the corresponding many-body interaction Hamiltonian, we employ the apparatus of spherical tensor operators \cite{Racah1959,Brink_1968,Ryabov_1999,Wigner_1959}. In view of the fact that a space is isotropic, the interaction Hamiltonian must be a scalar. Employing this fact and representation properties 
for the rotation group, we can write it in the following form: 
\begin{equation}\label{eq:Int}
V=
\sum_{i=0}^{2F}\sum_{m=-i}^{i}(-1)^{m}c_{i}T_{m}^{i}T_{-m}^{i}.
\end{equation}
Here $c_{i}$ is the coupling constant and $T^{i}_{m}$ is the  spherical tensor operator. By definition, this operator is irreducible and the upper and lower indices denote its rank and component, respectively (for a given rank $i$, $T^{i}_{m}$ has $2i+1$ components). Therefore, for any irreducible multipole operator of rank $i$, we can relate its components to the components of the same rank spherical tensor operator $T^{i}_{m}$. In particular, for atoms with $F=3/2$, the Hamiltonian given by Eq.~\eqref{eq:Int} has a clear physical meaning: it is a sum of spin, quadrupole, and octupole exchange interaction terms. 

Since all the components of a spherical tensor operator of rank $i$ can be expressed as a polynomial of degree $i$ in the components of the spin operators, the Hamiltonian \eqref{eq:Int} can be reduced to  \cite{Nagaev_1982} (see also Eq.~\eqref{eq:Int_mult})
\begin{equation}
    \label{eq:Ham_FF}
    V=\sum_{n=0}^{2F}\tilde{c}_{n}\,({\bf F}_{1}\cdot{\bf F}_{2})^{n}.  
\end{equation}
Besides that, there are other equivalent representations of the Hamiltonian with multipole exchange interaction. They allow one to write the Hamiltonian as a bilinear form of certain operators, like Steven's operators \cite{Stevens_1952,Rudowicz_2004} (see, e.g., \cite{Kosmachev2015} for $F=3/2$) or operators that provide a representation of the SU($2F+1$) group (e.g., Gell-Mann matrices for $F=1$; see \cite{Peletminskii_2020,Bulakhov_2021,Bulakhov_2022}).  All the mentioned ways of writing the Hamiltonian (including Eq.~\eqref{eq:Ham_FF}) are equivalent due to the fact that they involve a complete set of operators \cite{Kusunose_2008}. 

Nevertheless, in our subsequent study, we employ the apparatus of spherical tensor operators because it gives the transparent relation between the rotation group representation and theory of angular momentum. In particular, this relation allows us at once to formulate the Hamiltonian \eqref{eq:Int} as a scalar operator. In addition, the spherical tensor operator is the major constituent of the celebrated Wigner-Eckart theorem \cite{Wigner_1959,Eckart_1930}. By the way, according to this theorem, any pairwise interaction invariant under rotation is reduced to the form of Eq.~\eqref{eq:Int}. Just for this reason, the Hamiltonian of two hydrogen atoms interacting by means of Coloumb forces 
can be expressed in terms of the spin exchange interaction, 
$$
({\bf F}_{1}\cdot{\bf F}_{2})\propto \sum_{m=\pm 1,0}(-1)^{m}T^{1}_{m}T^{1}_{-m},
$$  
whereas the original Hamiltonian does not contain the concept of spin at all. 
In the context of our problem dealing with multipole exchange, the dipole-dipole interaction is equivalent to the quadrupole exchange interaction \cite{Racah1959}.

Since we are interested in the formulation of the many-body Hamiltonian, we address the second quantization method. According to its general rules for constructing physical quantities, we represent the Hamiltonian of interacting spin-3/2 atoms in the following form:
\begin{equation}\label{eq:TotHam}
    H=H_{0}+V,
\end{equation}
where $H_{0}$ includes the kinetic energy of atoms $\varepsilon_{\bf p}=p^{2}/2m$ and their coupling with a magnetic field,   
\begin{equation} \label{eq:Kin}    
   H_{0}=     \sum_{{\bf p}}a^{\dagger}_{{\bf p}\alpha}(\varepsilon_{\bf p}\delta_{\alpha\beta}-{\bf h}{\bf F}_{\alpha\beta})a_{{\bf p}\beta}.
\end{equation}
Here ${\bf h}=g\mu_{B}{\bf B}$ with $g$, $\mu_{B}$, and ${\bf B}$ being the Lande hyperfine factor, Bohr magneton, and external magnetic field, respectively.  
The interaction Hamiltonian, according to the structure of Eq.~(\ref{eq:Int}), reads
\begin{equation} \label{eq:Inter}
V=\frac1{2\mathcal{V}}
    \sum_{i=0}^{3}
    \sum_{m=-i}^{i}
    (-1)^{m}
    \sum_{{\bf p}_{1}\ldots{\bf p}_{4}}
    U^{(i)}({\bf p}_{1}-{\bf p}_{4})a^{\dagger}_{{\bf p}_{1}\alpha}a^{\dagger}_{{\bf p}_{2}\beta}
    (T^{i}_{m})_{\alpha\delta}(T^{i}_{-m})_{\beta\gamma}
    a^\pd_{{\bf p}_{3}\gamma}
    a^\pd_{{\bf p}_{4}\delta}\,\delta^\pd_{{\bf p}_{1}+{\bf p}_{2},\,{\bf p}_{3}+{\bf p}_{4}},
\end{equation}
where ${\cal V}$ is the volume of the system and $U^{(i)}({\bf p}_{1}-{\bf p}_{4})$ are the Fourier transforms of the energies corresponding to spin-independent ($i=0$) interaction as well as spin ($i=1$), quadrupole ($i=2$), and octupole ($i=3$) exchange interactions. The creation and annihilation operators of fermionic atoms meet the usual anticommutation relations, $\{a_{{\bf p}\alpha},a_{{\bf p}'\beta}\}=0$ and $\{a_{{\bf p}\alpha},a^{\dagger}_{{\bf p}'\beta}\}=\delta_{{\bf p}{\bf p}'}\delta_{\alpha\beta}$. In Eqs.~\eqref{eq:Kin}, \eqref{eq:Inter} and below, we assume the summation over the repeated indices, unless otherwise specified.

It is worth stressing that while describing the collision of two identical spin-$F$ atoms within the scattering-length approximation, only the scattering channels with even total spin ${\cal F}$ are open regardless of the Fermi-Dirac or Bose-Einstein statistics \cite{UedaPhysRep}. Therefore, in this case, not all multipole moments are included in the description of the exchange interaction (for details, see \ref{app:Projection_Op}).

\section{Kinetic equation for fermionic atoms in the weak interaction approximation}

Remind that the temporal evolution of the system essentially depends on hierarchy of two typical time scales. The first one is the duration of a collision event $\tau_{0}=r_{0}/v$, where $r_{0}$ is the interaction range and $v$ is the average particle velocity. The second characteristic value is the relaxation time $\tau_{r}=l/v$, where $l=(n\sigma)^{-1}$ is the mean free path, $n$ is the particle density and $\sigma\propto r_{0}^{2}$ is the particle scattering cross-section. Therefore, $\tau_{0}/\tau_{r}=(r_{0}/a)^{3}$, where $a$ is the average distance between particles. For gases, where $r_{0}\ll a$, one finds a characteristic separation of time scales, i.e., $\tau_{r}\gg\tau_{0}$. It is this separation of time scales gives rise to a kinetic stage of evolution. However, a kinetic stage is absent in dense systems for which $r_{0}\sim a$.  In such systems, where a local equilibrium is established rapidly over the time $\tau_{r}$, the only hydrodynamic stage of evolution remains. 

The reduced description method of quantum many-body systems \cite{Yatsenko1968,AkhPel} based on the temporal hierarchy is a powerful tool for deriving kinetic or hydrodynamic equations. 
The main ingredient of this method is a coarse-grained statistical operator that
depends only on a number of so-called reduced description parameters. These parameters serving as master variables determine the
evolution of the system on a coarse-grained time scale. In particular, for quantum systems with a weak interaction (atomic gases), the single-particle density matrix $f_{i_{1},i_{2}}$ is chosen as the master variable on time scales $t\gg\tau_{0}$. From
a perturbative expansion of the corresponding coarse-grained statistical operator, one can arrive at the following kinetic equation in operator form \cite{AkhPel}:
\begin{equation}\label{eq:KinEq}
    \frac{\partial f}{\partial t} + \frac{i}{\hbar}\left[\varepsilon,\, f\right] = L^{(2)}(f),
\end{equation}
where
\begin{equation}\label{eq:QuasEn}
    \varepsilon_{i_1,i_2}=\varepsilon_{i_1}\delta_{i_1,i_2}+\frac{1}{\cal V}\sum_{i_1',i_2'}\Phi(i_1,i_1';i_2',i_2) f_{i_2',i_1'}
\end{equation}
and the quantities $\varepsilon_{i_1}$ and $\Phi(i_{1},{i}_2;i_{3},i_{4})$ determine the microscopic Hamiltonian,
\begin{equation}\label{eq:QuasHam}
    H=\sum_{i}\varepsilon_{i}a_{i}^{\dagger}a_{i}^{\pd}
    +
    \frac{1}{4\mathcal{V}}
    \sum_{i_{1}\ldots i_{4}}\Phi(i_1,i_2;i_3,i_4)a^{\dagger}_{i_1}a^{\dagger}_{i_2}a_{i_3}^{\pd}a_{i_4}^{\pd}.
\end{equation}
Obviously, for fermionic atoms, the interaction amplitude has the following symmetry properties:
\begin{equation}
    \Phi(i_1,i_2;i_3,i_4)=-\Phi(i_1,i_2;i_4,i_3)=-\Phi(i_2,i_1;i_3,i_4)=\Phi(i_2,i_1;i_4,i_3).
\end{equation}
Index $i$ denotes the whole set of quantum numbers that define the individual particle state. In our case, the state of an atom is given by its momentum ${\bf p}$ and spin projection $\alpha$, so that $i=\{{\bf p},\alpha\}$.
The right-hand side of Eq.~(\ref{eq:KinEq}) represents a collision integral $L^{(2)}(f)$ in the second order in interaction. For the sake of brevity, we do not write its explicit form (for details, see Ref.~\cite{AkhPel}) since below we are interested in solving the kinetic equation in the collisionless approximation. 
Note that the left-hand side of kinetic equation is determined by the quantity $\varepsilon_{1,2}$ which, according to Eq.~(\ref{eq:QuasEn}), depends both on interaction amplitude and single-particle density matrix. Therefore, this quantity, being a modified (quasi)particle energy, involves the mean-field effects. 
The kinetic equation (\ref{eq:KinEq}) preserves its general structure also for bosonic atoms and the difference is revealed  
in the explicit form of the collision integral \cite{AkhPel}. 
It is applicable to describe both homogeneous and inhomogeneous quantum gases in the normal state with no broken symmetries. For superfluid gases, the number of reduced description parameters (or master variables) is not limited to the single-particle density matrix and the derivation of the corresponding kinetic equation should be substantially modified (see Refs.~\cite{Shchelokov1977,Kirkpatrick1985,Walser1999,Reichl2013,Reichl2019} and Ref.~\cite{Galaiko1972} for Bose and Fermi gases, respectively).

To proceed further, we introduce the Wigner distribution function,
\begin{multline}\label{eq:Wigner_f}
    f_{\alpha\beta}
    ({\bf x},{\bf p})
    =\sum_{\bf k}e^{-i{\bf kx}}f_{\alpha\beta}\left({\bf p}-\frac{{\bf k}}{2},{\bf p}+\frac{{\bf k}}{2}\right)=
    \\
    \frac{\mathcal{V}}{(2\pi\hbar)^3}
    \int
    d{\bf k}
    \,
    e^{-i{\bf kx}}
    f_{\alpha\beta}\left({\bf p}-\frac{{\bf k}}{2},{\bf p}+\frac{{\bf k}}{2}\right).
\end{multline}
In a similar manner, one can define the particle energy  dependent on space coordinate and momentum,
\begin{multline}\label{eq:Wigner_e}
    \varepsilon_{\alpha\beta}
    ({\bf x},{\bf p})
    =\sum_{\bf k}e^{-i{\bf kx}}\varepsilon_{\alpha\beta}\left({\bf p}-\frac{{\bf k}}{2},{\bf p}+\frac{{\bf k}}{2}\right)
    =\\
    \frac{\mathcal{V}}{(2\pi\hbar)^3}
    \int
    d{\bf k}
    \,
    e^{-i{\bf kx}}
    \varepsilon_{\alpha\beta}\left({\bf p}-\frac{{\bf k}}{2},{\bf p}+\frac{{\bf k}}{2}\right).
\end{multline}
Then, we can rewrite the kinetic equation \eqref{eq:KinEq} for weakly inhomogeneous states. Employing the mathematical procedure \cite{AkhPel}, one obtains  
\begin{gather}
    \frac{\partial}{\partial t}
    f_{\alpha\beta}
    ({\bf x},{\bf p})
    +
    \frac{i}{\hbar}
    \left[
        \varepsilon
        ({\bf x},{\bf p})
        ,
        f
        ({\bf x},{\bf p})
    \right]
    _{\alpha\beta}
    \nonumber\\+
    \frac12
    \left\{
        \frac{\partial\varepsilon
        ({\bf x},{\bf p})}{\partial {\bf p}}
        ,
        \frac{\partial f
        ({\bf x},{\bf p})}{\partial {\bf x}}
    \right\}_{\alpha\beta}
    -
    \frac12
    \left\{
        \frac{\partial\varepsilon
        ({\bf x},{\bf p})}{\partial {\bf x}}
        ,
        \frac{\partial f
        ({\bf x},{\bf p})}{\partial {\bf p}}
    \right\}_{\alpha\beta}
    =0,
    \label{eq:KinEqX}
\end{gather}
where the curly brackets denote anticommutator. It is worth stressing that the kinetic equation \eqref{eq:KinEqX} is valid for systems, where the characteristic scales of spatial inhomogeneities are large compared to the interaction range $r_{0}$ and to the De Broglie wavelength $\lambda=\hbar/mv$. Obviously, it is applicable to describe quantum gases of atoms with arbitrary spin $F$. Finally, for spinless atoms, the kinetic equation \eqref{eq:KinEqX} is formally  reduced to the usual classical form.

In order to adapt the derived kinetic equation to describe the quantum gas of interacting spin 3/2 atoms, we need to express the energy $\varepsilon_{\alpha\beta}({\bf x},{\bf p})$ in terms of the quantities entering the microscopic Hamiltonian given by Eqs.~\eqref{eq:TotHam}--\eqref{eq:Inter}. Therefore, we need to symmetrize Eq.~\eqref{eq:Inter}:
\begin{gather}
V=\frac{1}{8{\cal V}} \sum_{i=0}^{3}
    \sum_{m=-i}^{i}(-1)^{m}
    \sum_{{\bf p}_{1}\ldots{\bf p}_{4}}a^{\dagger}_{{\bf p}_{1}\alpha}a^{\dagger}_{{\bf p}_{2}\beta}  a^\pd_{{\bf p}_{3}\gamma}
    a^\pd_{{\bf p}_{4}\delta}\,\delta^\pd_{{\bf p}_{1}+{\bf p}_{2},\,{\bf p}_{3}+{\bf p}_{4}} \nonumber \\
    \times
    [U^{(i)}({\bf p}_{1}-{\bf p}_{4}) (T^{i}_{m})_{\alpha\delta}(T^{i}_{-m})_{\beta\gamma}+U^{(i)}({\bf p}_{2}-{\bf p}_{3}) (T^{i}_{m})_{\beta\gamma}(T^{i}_{-m})_{\alpha\delta} \nonumber
    \\
        -U^{(i)}({\bf p}_{2}-{\bf p}_{4}) (T^{i}_{m})_{\beta\delta}(T^{i}_{-m})_{\alpha\gamma}-U^{(i)}({\bf p}_{1}-{\bf p}_{3}) (T^{i}_{m})_{\alpha\gamma}(T^{i}_{-m})_{\beta\delta}]. \label{eq:SymmHam}
\end{gather}
The comparison of Eq.~\eqref{eq:SymmHam} to  Eq.~\eqref{eq:QuasHam} in which $\Phi(i_1,i_2;i_3,i_4)\equiv \Phi_{\alpha\beta\gamma\delta}({\bf p}_{1},{\bf p}_{2},{\bf p}_{3},{\bf p}_{4})$ gives
\begin{gather*}
    \Phi_{\alpha\beta\gamma\delta}({\bf p}_{1},{\bf p}_{2},{\bf p}_{3},{\bf p}_{4})=\frac{1}{2}\sum_{i=0}^{3}
    \sum_{m=-i}^{i}(-1)^{m} \delta^\pd_{{\bf p}_{1}+{\bf p}_{2},\,{\bf p}_{3}+{\bf p}_{4}} \nonumber \\
    \times
    [U^{(i)}({\bf p}_{1}-{\bf p}_{4}) (T^{i}_{m})_{\alpha\delta}(T^{i}_{-m})_{\beta\gamma}+U^{(i)}({\bf p}_{2}-{\bf p}_{3}) (T^{i}_{m})_{\beta\gamma}(T^{i}_{-m})_{\alpha\delta} \nonumber
    \\
        -U^{(i)}({\bf p}_{2}-{\bf p}_{4}) (T^{i}_{m})_{\beta\delta}(T^{i}_{-m})_{\alpha\gamma}-U^{(i)}({\bf p}_{1}-{\bf p}_{3}) (T^{i}_{m})_{\alpha\gamma}(T^{i}_{-m})_{\beta\delta}].
\end{gather*}
Next, substituting this relation into Eq.~(\ref{eq:QuasEn}) and changing to the function $f_{\alpha\beta}({\bf x},{\bf p})$, $\varepsilon_{\alpha\beta}({\bf x},{\bf p})$ according to Eqs.~(\ref{eq:Wigner_f}) and (\ref{eq:Wigner_e}), one obtains
\begin{gather}
\varepsilon_{\alpha\beta}({\bf x},{\bf p})=\varepsilon_{\bf p}\delta_{\alpha\beta}
    -
    ({\bf hF})_{\alpha\beta} \nonumber  
    \\
    +\frac{1}{\cal V}\sum_{i=0}^{3}\sum_{m=-i}^{i}(-1)^{m}\sum_{{\bf p}'}\int d{\bf x}'\,\mathcal{U}^{(i)}({\bf x}-{\bf x}')(T^{i}_{m})_{\alpha\beta}(T^{i}_{-m})_{\delta\gamma}f_{\gamma\delta}({\bf x}',{\bf p}') \nonumber \\
-\frac{1}{\cal V}\sum_{i=0}^{3}\sum_{m=-i}^{i}(-1)^{m}\sum_{{\bf p}'}U^{(i)}({\bf p}-{\bf p}')(T^{i}_{m})_{\alpha\gamma}f_{\gamma\delta}({\bf x},{\bf p}')(T^{i}_{-m})_{\delta\beta}, \label{eq:QuasEnX}
\end{gather}
where the first two terms come from the operator $H_{0}$ (see Eq.~(\ref{eq:Kin})) and 
$$
U^{(i)}({\bf p})=\int d{\bf x}\, \mathcal{U}^{(i)}({\bf x}) e^{-\frac{i}{\hbar}({\bf px})}.
$$ 
Equation \eqref{eq:QuasEnX} represents the general expression for the particle energy, which includes the mean-field effects and multipole degrees of freedom. 
However, we can assume that the interaction range of potential is negligibly small compared to the scale of spatial inhomogeneities (the distance over which the distribution function changes).
In this case the function $f_{\gamma\delta}({\bf x}',{\bf p}')$ can be taken out of the integral at the point ${\bf x}$ and Eq.~\eqref{eq:QuasEnX} takes a more simple form,
\begin{gather}
\varepsilon_{\alpha\beta}({\bf x},{\bf p})=\varepsilon_{\bf p}\delta_{\alpha\beta}
    -
    ({\bf hF})_{\alpha\beta} \nonumber  
    \\
    +\frac{1}{\cal V}\sum_{i=0}^{3}\sum_{m=-i}^{i}(-1)^{m}\sum_{{\bf p}'} U^{(i)}(0)(T^{i}_{m})_{\alpha\beta}(T^{i}_{-m})_{\delta\gamma}f_{\gamma\delta}({\bf x},{\bf p}') \nonumber \\
-\frac{1}{\cal V}\sum_{i=0}^{3}\sum_{m=-i}^{i}(-1)^{m}\sum_{{\bf p}'}U^{(i)}({\bf p}-{\bf p}')(T^{i}_{m})_{\alpha\gamma}f_{\gamma\delta}({\bf x},{\bf p}')(T^{i}_{-m})_{\delta\beta}.\label{eq:QuasEnXSimp}
\end{gather}
Below, we are interested in solving the kinetic equation (\ref{eq:KinEqX}) with the given particle energy $\varepsilon_{\alpha\beta}({\bf x},{\bf p})$ determined by Eq.~\eqref{eq:QuasEnXSimp}. 
Moreover, we shall assume that ${\bf h}=(0,0,h)$ and, consequently, $({\bf hF})_{\alpha\beta}=(hF^z)_{\alpha\beta}$.

\section{Linearized kinetic equation}

In this section, we study the propagation of small oscillations known as zero sound in a quantum Fermi gas of interacting spin-3/2 atoms. To this end, we employ the kinetic equation \eqref{eq:KinEqX} supplemented by Eq.~(\ref{eq:QuasEnXSimp}) for the particle energy. The collisionless approximation is justified by the fact that the collision integral $L^{(2)}(f)\sim\tau_r^{-1}$, where $\tau_r$ is the relaxation time of the distribution function. Therefore, we can neglect $L^{(2)}(f)$ when studying the high-frequency oscillations with $\omega\tau_r\gg1$.  

We are interested in solving the kinetic equation perturbatively, by its linearization close to equilibrium state which is given by the diagonal single-particle density matrix,
\begin{equation}\label{eq:EqDistr}
f_{\alpha\beta}({\bf p})=f^{{[\alpha]}}_{\bf p}\delta_{\alpha\beta},
\end{equation}
where 
$f^{[\alpha]}_{\bf p}$ are the Fermi-Dirac distribution functions corresponding to four eigenvalues of the operator $F^{z}$,
\begin{gather*}
f^{[1]}_{\bf p}=\left(e^{\beta(\varepsilon_{\bf p}-\mu-3h/2)}+1\right)^{-1}, \quad f^{[2]}_{\bf p}=\left(e^{\beta(\varepsilon_{\bf p}-\mu-h/2)}+1\right)^{-1}, \\
f^{[3]}_{\bf p}=\left(e^{\beta(\varepsilon_{\bf p}-\mu+h/2)}+1\right)^{-1}, \quad
f^{[4]}_{\bf p}=\left(e^{\beta(\varepsilon_{\bf p}-\mu+3h/2)}+1\right)^{-1},
\end{gather*}
and $\mu$ is the chemical potential. These functions are combined to be
\begin{equation}\label{eq:GenDF}
f^{[\alpha]}_{\bf p}=\left(e^{\beta(\varepsilon_{\bf p}-\tilde{\mu}^{[\alpha]})}+1\right)^{-1}, \quad \tilde{\mu}^{[\alpha]}=\mu+h\left(\frac{5}{2}-\alpha\right).
\end{equation}

To proceed further, it is convenient to decompose $f_{\alpha\beta}({\bf x},{\bf p})$ and $\varepsilon_{\alpha\beta}({\bf x},{\bf p})$ into complete set of spherical tensor operators,  
\begin{equation} \label{eq:Decomp}
    f_{\alpha\beta}({\bf x},{\bf p})=
    \sum_{i=0}^{3}\sum_{m=-i}^{i}
    f^{i}_{m}({\bf x},{\bf p})(T^{i}_{m})_{\alpha\beta}
    ,\quad
    \varepsilon_{\alpha\beta}({\bf x},{\bf p})=
    \sum_{i=0}^{3}\sum_{m=-i}^{i}
    \varepsilon^{i}_{m}({\bf x},{\bf p})(T^{i}_{m})_{\alpha\beta}.
\end{equation}
The decomposition coefficients are found by using the normalization condition given by Eq.~\eqref{eq:Norm},
\begin{equation} \label{eq:Coeff}
  f^{i}_{m}({\bf x},{\bf p})=(-1)^{m}f_{\alpha\beta}({\bf x},{\bf p})(T^{i}_{-m})_{\beta\alpha},
  \quad
\varepsilon^{i}_{m}({\bf x},{\bf p})=(-1)^{m}\varepsilon_{\alpha\beta}({\bf x},{\bf p})(T^{i}_{-m})_{\beta\alpha}.  
\end{equation}
Note that it is the coefficients $f^{i}_{m}({\bf x},{\bf p})$ that 
are related to different physical quantities such as density ($i=0$), magnetization ($i=1$), as well as quadrupole ($i=2$) and octupole ($i=3$) moments. 
The index $m$ enumerates the components of each quantity and takes $2i+1$ values.

The conditions for the applicability of the perturbative approach under consideration can be formulated as follows:  
\begin{gather}
    f^{i}_{m}({\bf x},{\bf p})
    \approx
    f^{i}_{m}({\bf p})
    +
    \tilde{f}^{i}_{m}({\bf x},{\bf p})
    ,\quad
    |f^{i}_{m}({\bf p})|
    \gg
    |\tilde{f}^{i}_{m}({\bf x},{\bf p})|
    , \nonumber \\
    \varepsilon^{i}_{m}({\bf x},{\bf p})
    \approx
    \varepsilon^{i}_{m}({\bf p})
    +
    \tilde{\varepsilon}^{i}_{m}({\bf x},{\bf p})
    ,\quad
    |\varepsilon^{i}_{m}({\bf p})|
    \gg
    |\tilde{\varepsilon}^{i}_{m}({\bf x},{\bf p})|,
    \label{eq:PertCond}
\end{gather}
where the tilde denotes the deviation of the corresponding quantity from its equilibrium value, which does not depend on the space coordinate. According to Eq.~\eqref{eq:QuasEnXSimp}, $\varepsilon_{m}^{i}({\bf p})$ and $\tilde{\varepsilon}_{m}^{i}({\bf x},{\bf p})$ are induced by the equilibrium and perturbed values of the single-particle density matrix, respectively.
Noting that $f^{i}_{m}({\bf p})=(-1)^{m}f_{\alpha\beta}({\bf p})(T^{i}_{-m})_{\beta\alpha}$ and employing Eq.~\eqref{eq:EqDistr}, 
 one obtains $f^{i}_{m}({\bf p})=(-1)^{m}f^{[\alpha]}_{\bf p}(T_{-m}^{i})_{\alpha\alpha}$. Since all operators $T^{i}_{m}$, except $T^{i}_{0}$, have only zeros in their diagonals (see \ref{app:STO}), we conclude that only the terms with $m=0$ contribute to the decomposition of $f_{\alpha\beta}({\bf x},{\bf p})$,
 \begin{equation} \label{eq:DiagDM}
 f_{\alpha\beta}({\bf x},{\bf p})=\sum_{i=0}^{3}f_{0}^{i}({\bf x},{\bf p})(T^{i}_{0})_{\alpha\beta}.
\end{equation}
Consequently, $f_{\alpha\beta}({\bf x},{\bf p})$ is a diagonal matrix due to the diagonal structure of all $(T^{i}_{0})_{\alpha\beta}$, see \ref{app:STO}. Employing this fact, one can easily prove that $\varepsilon_{\alpha\beta}({\bf x},{\bf p})$ determined by Eq.~\eqref{eq:QuasEnXSimp} represents also a diagonal operator (the diagonal structure of the first two terms is obvious). Indeed, following to  Eqs.~\eqref{eq:DiagDM} and \eqref{eq:Norm}, it is easy to see that only the operators $(T^{i}_{0})_{\alpha\beta}$ contribute to the third term of Eq.~\eqref{eq:QuasEnXSimp} making it diagonal.
As for the fourth term, it contains the product $(T^{i}_{m})_{\alpha\gamma}(T^{i}_{-m})_{\gamma\beta}$, which represents a diagonal operator for any $i$ and respective $m$. Hence, according to the second formula in \eqref{eq:Decomp}, $\varepsilon_{\alpha\beta}({\bf x},{\bf p})$ is decomposed into the diagonal operators $(T^{i}_{0})_{\alpha\beta}$, 
$$
\varepsilon_{\alpha\beta}({\bf x},{\bf p})=\sum_{i=0}^{3}\varepsilon_{0}^{i}({\bf x},{\bf p})(T^{i}_{0})_{\alpha\beta}.
$$
The equilibrium $\varepsilon_{\alpha\beta}({\bf p})$ and perturbed $\tilde{\varepsilon}_{\alpha\beta}({\bf x},{\bf p})$ parts of the particle energy are found from Eqs.~\eqref{eq:QuasEnXSimp},
\begin{gather}
\varepsilon_{\alpha\beta}({\bf p})=\varepsilon_{\bf p}\delta_{\alpha\beta}
    -
    hF^{z}_{\alpha\beta}
    +\frac{1}{\cal V}\sum_{i=0}^{3}\sum_{{\bf p}'} U^{(i)}(0)(T^{i}_{0})_{\alpha\beta}(T^{i}_{0})_{\gamma\gamma}f^{[\gamma]}_{{\bf p}'} \nonumber \\
-\frac{1}{\cal V}\sum_{i=0}^{3}\sum_{m=-i}^{i}(-1)^{m}\sum_{{\bf p}'}U^{(i)}({\bf p}-{\bf p}')(T^{i}_{m})_{\alpha\gamma}(T^{i}_{-m})_{\gamma\beta}f^{[\gamma]}_{{\bf p}'} \label{eq:QuasEnEq}
\end{gather}
and
\begin{gather}
\tilde{\varepsilon}_{\alpha\beta}({\bf x},{\bf p})
=
    \frac{1}{\cal V}\sum_{i=0}^{3}\sum_{{\bf p}'} U^{(i)}(0)(T^{i}_{0})_{\alpha\beta}(T^{i}_{0})_{\gamma\gamma}\tilde{f}^{[\gamma]}({\bf x},{\bf p}') 
    \nonumber \\
-\frac{1}{\cal V}\sum_{i=0}^{3}\sum_{m=-i}^{i}(-1)^{m}\sum_{{\bf p}'}U^{(i)}({\bf p}-{\bf p}')(T^{i}_{m})_{\alpha\gamma}(T^{i}_{-m})_{\gamma\beta}\tilde{f}^{[\gamma]}({\bf x},{\bf p}'). \label{eq:QuasEnPert}
\end{gather}
Here and below, no summation is assumed over the index in square brackets (it is neither a vector nor a tensor index).

Let us return to the kinetic equation \eqref{eq:KinEqX}. Taking into account the diagonal structure of the matrices $f_{\alpha\beta}({\bf x},{\bf p})$ and $\varepsilon_{\alpha\beta}({\bf x},{\bf p})$ and making the substitution  
$f_{\alpha\beta}({\bf x},{\bf p})=(f^{[\alpha]}_{\bf p}+\tilde{f}^{[\alpha]}({\bf x},{\bf p}))\delta_{\alpha\beta}$, we can recast it in the following linearized form: 
\begin{equation}
    \frac{\partial \tilde{f}^{[\alpha]}
    ({\bf x},{\bf p})}{\partial t}
    \delta_{\alpha\beta}
        +
    \frac{\partial\varepsilon_{\alpha\beta}
    ({\bf p})}{\partial {\bf p}}
    \frac{\partial\tilde{f}^{[\alpha]}
    ({\bf x},{\bf p})}{\partial {\bf x}}
    -
    \frac{\partial\tilde{\varepsilon}_{\alpha\beta}({\bf x},{\bf p})}{\partial {\bf x}}
    \frac{\partial f^{[\alpha]}_{\bf p}}{\partial {\bf p}}
    =0
    .
    \label{eq:KinEq1}
\end{equation}
We seek the solution of Eq.~\eqref{eq:KinEq1} in the form of the Fourier transform for the distribution function,   
\begin{equation*}
    \tilde{f}^{[\alpha]}
    ({\bf x},{\bf p},t)
    =
    \frac{1}{(2\pi)^{4}}
    \int
    d^3{\bf k}
    \,
    d\omega\:
    g^{[\alpha]}({\bf k},\omega,{\bf p})
    e^{i{\bf kx}-i\omega t}
    .
\end{equation*}
This trick allows us to get rid of the differentiation:
\begin{gather}
    \left(
            \omega
             -
            {\bf k}
            \frac{\bf p}{m}
            +
            {\bf k}{\bf A}^{[\alpha]}({\bf p})
    \right)
    g^{[\alpha]}({\bf k},\omega,{\bf p})
    \delta_{\alpha\beta}
    +
    {\bf k}\frac{\bf p}{m}
    \frac{
        \partial f^{[\alpha]}_{\bf p}
        }{
        \partial \varepsilon_{\bf p}
        }
    \frac{1}{\mathcal{V}}
    \sum_{i=0}^3
    \sum_{{\bf p}^{\prime}}
    g^{[\gamma]}({\bf k},\omega,{\bf p}')
    \nonumber \\
    \times
    \Bigg(
        U^{(i)}(0)
       (T^{i}_{0})_{\gamma\gamma}
       (T^{i}_{0})_{\alpha\beta}
    \Bigg.
    -
    \left.
    \sum_{m=-i}^{i}(-1)^{m}
    U^{(i)}({\bf p}-{\bf p}^{\prime})    
    (T^{i}_{m})_{\alpha\gamma}
    (T^{i}_{-m})_{\gamma\beta}
    \right)
    =
    0
    ,
    \label{eq:KinEqg}
\end{gather}
where
\begin{equation}\label{eq:FuncA}
    {\bf A}_{\alpha\beta}({\bf p})
    \equiv
    \frac{1}{\mathcal{V}}
    \sum_{i=0}^{3}
    \sum_{m=-i}^{i}(-1)^{m}
    \dfrac{\partial}{{\partial\bf p}}
    \sum_{{\bf p}'} 
    U^{(i)}({\bf p}-{\bf p}')
    (T^{i}_{m})_{\alpha\gamma}
    f^{[\gamma]}_{{\bf p}'}
    (T^{i}_{-m})_{\gamma\beta}
    =
    {\bf A}^{[\alpha]}({\bf p})
    \delta_{\alpha\beta}
    .
\end{equation}
Taking into account the structure of the coupled equations \eqref{eq:KinEqg}, it is appropriate to make the following substitution:
\begin{equation*}
    g^{[\alpha]}({\bf k},\omega,{\bf p})=
    \frac{\partial f^{[\alpha]}
    _{\bf p}}{\partial \varepsilon_{\bf p}}
    {\bf k}
    \frac{\bf p}{m}
    \left(
            \omega
             -
            {\bf k}
            \frac{\bf p}{m}
            +
            {\bf k}{\bf A}^{[\alpha]}({\bf p})
    \right)^{-1}
    y^{[\alpha]}(p)
    ,
\end{equation*}
where the introduced function $y^{[\alpha]}(p)$ depends on the modulus of the vector ${\bf p}$. Thus, Eqs.~\eqref{eq:KinEqg} read, 
\begin{gather}
    y^{[\alpha]}(p)
    \delta_{\alpha\beta}
    +
    \frac{1}{\mathcal{V}}
    \sum_{i=0}^3
    \sum_{{\bf p}^{\prime}}
    \frac{\partial f^{[\gamma]}
    _{{\bf p}'}}{\partial \varepsilon_{{\bf p}'}}
    {\bf k}
    \frac{{\bf p}'}{m}
    \left(
            \omega
             -
            {\bf k}
            \frac{{\bf p}'}{m}
            +
            {\bf k}{\bf A}^{[\gamma]}({\bf p}')
    \right)^{-1}
    y^{[\gamma]}(p')
    \nonumber \\
    \times
    \Bigg(
        U^{(i)}(0)
       (T^{i}_{0})_{\gamma\gamma}
       (T^{i}_{0})_{\alpha\beta}
    \Bigg.
    -
    \left.
    \sum_{m=-i}^{i}(-1)^{m}
    U^{(i)}({\bf p}-{\bf p}^{\prime})    
    (T^{i}_{m})_{\alpha\gamma}
    (T^{i}_{-m})_{\gamma\beta}
    \right)
    =
    0.
     \label{eq:KinEqy}
\end{gather}
The coupled equations \eqref{eq:KinEqy} provide the platform for analysing the dispersion relations of zero sound oscillations.

\section{Dispersion equation and zero sound}

As we mentioned, the specific feature of zero-sound associated with fluctuation of the fermionic distribution function is that it propagates even at zero temperature. In this temperature limit, the distribution function $f^{[\alpha]}_{\bf p}$ given by Eq.~\eqref{eq:GenDF} takes the form
\begin{equation}\label{eq:DFZero}
    f^{[\alpha]}_{\bf p}
    =
    \Theta
    (
        \varepsilon_{\rm F}^{[\alpha]}
        -
        \varepsilon_{\bf p}
    )
    ,\quad
    \frac{\partial f^{[\alpha]}_{\bf p}}{\partial \varepsilon_{\bf p}}
    =
    -\delta
    (
        \varepsilon_{\rm F}^{[\alpha]}
        -
        \varepsilon_{\bf p}
    )
    \end{equation}
where $\Theta(x)$ is the Heaviside step function and 
\begin{equation}\label{eq:FermiEn}
\varepsilon_{\rm F}^{[\alpha]}(h)=\varepsilon_{\rm F}(h)+h\left(\dfrac{5}{2}-\alpha\right).
\end{equation}
Calculating the total number of spin-3/2 atoms with the distribution function \eqref{eq:DFZero}, we have
\begin{equation} \label{eq:EnF}
    \sum_{\alpha=1}^{4}
    \Theta(\varepsilon_{\rm F}^{[\alpha]})
    [\varepsilon_{\rm F}^{[\alpha]}]^{3/2}
    =4[\varepsilon_{\rm F}(0)]^{3/2}, 
    \quad
    \varepsilon_{\rm F}(0)=\dfrac{\hbar^2}{2^{5/3}m}\left(3\pi^2n\right)^{2/3}
    ,
\end{equation}
where $n$ is the total atomic density. 
At zero temperature limit, the system of coupled integral equations \eqref{eq:KinEqy} with respect to $y^{[\alpha]}(p)$ makes sense only for $p=p_{\rm F}^{[\alpha]}$, where $p_{\rm F}^{[\alpha]}=(2m\varepsilon_{\rm F}^{[\alpha]})^{1/2}$, otherwise the system decouples. Furthermore, changing the summation by integration over ${\bf p}'$ and employing the second relation in Eqs.~\eqref{eq:DFZero}, we obtain the system of linear equations with respect to $y_{\rm F}^{[\alpha]}$ instead of integral ones, 
\begin{gather}
    y^{[\alpha]}_{\rm F}
    \delta_{\alpha\beta}
    -
    \frac{m^{3/2}\sqrt{2\varepsilon_{\rm F}(0)}}{(2\pi\hbar)^3}
    \sum_{i=0}^3
    \int\limits_{0}^{2\pi}
    d\phi
    \int\limits_{0}^{\pi}
    \sin\theta d\theta\,
    \dfrac{
        \epsilon_{\rm F}^{[\gamma]}
        \cdot
        \Theta
        (\epsilon_{\rm F}^{[\gamma]})
        \cdot
        \cos\theta
    }{
        w
        -
        \left[
            \sqrt{\epsilon_{\rm F}^{[\gamma]}}
            -
            a^{[\gamma]}_{\rm F}
        \right]
        \cos\theta
    }
    \ 
    y^{[\gamma]}_{\rm F}
    \nonumber\\
    \times
    \left(
        U^{(i)}(0)
        (T^{i}_{0})_{\gamma\gamma}
        (T^{i}_{0})_{\alpha\beta}
        -
        \sum_{m=-i}^{i}(-1)^{m}
        U^{(i)}(\epsilon_{\rm F}^{[\alpha]},\epsilon_{\rm F}^{[\gamma]},\cos\chi(\theta,\phi,\kappa))
        (T^{i}_{m})_{\alpha\gamma}
        (T^{i}_{-m})_{\gamma\beta}
    \right)
    =
    0
    ,
    \label{eq:KinEqy1}
\end{gather}
where $w=\dfrac{\omega}{k}\sqrt{\dfrac{m}{2\varepsilon_{\rm F}(0)}}$ is the dimensionless speed of zero sound and the quantities
$$
y_{\rm F}^{[\alpha]}=y^{[\alpha]}(\epsilon_{\rm F}^{[\alpha]}), 
\quad
a_{\rm F}^{[\alpha]}=\sqrt{\dfrac{m}{2\varepsilon_{\rm F}(0)}}|{\bf A}^{[\alpha]}(\epsilon_{\rm F}^{[\alpha]})|,
$$ 
\begin{equation*}
U^{(i)}(\epsilon_{\rm F}^{[\alpha]},\epsilon_{\rm F}^{[\gamma]},\cos\chi(\theta,\phi,\kappa))\equiv U^{(i)}(\epsilon_{\rm F}^{[\alpha]}+\epsilon_{\rm F}^{[\gamma]}-2\sqrt{\epsilon_{\rm F}^{[\alpha]}\epsilon_{\rm F}^{[\gamma]}}\cos\chi(\theta,\phi,\kappa))
\end{equation*}
are considered as functions of dimensionless Fermi-energy
$\epsilon_{\rm F}^{[\alpha]}=\dfrac{\varepsilon_{\rm F}^{[\alpha]}}{\varepsilon_{\rm F}(0)}$. Three angles entering the above equation are defined as follows:
 $\kappa=\angle({\bf p},{\bf k})$, $\theta=\angle({\bf p}',{\bf k})$ and $\chi=\angle({\bf p},{\bf p}')$. It should be noted that they are  related by 
 \begin{equation*}
    \cos\chi(\theta,\phi,\kappa)
    =
    \sin\kappa
    \cos\phi
    \sin\theta
    +
    \cos\kappa
    \cos\theta.
\end{equation*}

The system of linear equations has a non-trivial solution provided that the respective determinant is equal to zero. In this regard, we note that the kinetic equation itself underlying this study is valid in the linear order in interaction. Therefore, while calculating the determinant, we must also keep only the linear terms with respect to this parameter. It is easy to see that such terms appear only from the entries of its main diagonal. Therefore, the resulting dispersion (characteristic) equation reads,
\begin{equation}\label{eq:DispEq}
 Y(w,h,\kappa)\approx 0,   
\end{equation}
where
\begin{gather}
    Y(w,h,\kappa)=1
    -
    \frac{m^{3/2}\sqrt{2\varepsilon_{\rm F}(0)}}{(2\pi\hbar)^3}
    \sum_{i=0}^3
    \int\limits_{0}^{2\pi}
    d\phi
    \int\limits_{0}^{\pi}
    \sin\theta d\theta\,
    \dfrac{
        \epsilon_{\rm F}^{[\alpha]}
        \cdot
        \Theta
        (\epsilon_{\rm F}^{[\alpha]})
        \cdot
        \cos\theta
    }{
        w
        -
        \left[
            \sqrt{\epsilon_{\rm F}^{[\alpha]}}
            -
            a^{[\alpha]}_{\rm F}
        \right]
        \cdot
        \cos\theta
    }
    \nonumber\\
    \times
    \left(
        U^{(i)}(0)
        -
        U^{(i)}(2\epsilon_{\rm F}^{[\alpha]}[1-\cos\chi(\theta,\phi,\kappa)])
    \right)
    ((T^{i}_{0})_{\alpha\alpha})^2
    . \label{eq:DispFun}
\end{gather}
This equation has no solution for contact interaction, i.e., $U^{(i)}(2\epsilon_{\rm F}^{[\alpha]}[1-\cos\chi(\theta,\phi,\kappa)])=U^{(i)}(0)$. Hence, one can claim that the phenomenon of zero sound in the system is due to finite-range interaction, even arbitrary small. As was claimed, in this case we must involve the interaction of multipole degrees of freedom. To proceed to further study of Eqs.~\eqref{eq:DispEq} and \eqref{eq:DispFun}, we need to specify the interaction potential. As such, we take the model potential of semitransparent spheres \cite{Bulakhov2018},
\begin{equation}\label{eq:potenr}
	\mathcal{U}^{(i)}({\bf x}-{\bf x}')
	=
	\left\{
	\begin{array}{cc}
		\dfrac{3g^{(i)}}{4\pi \left(r^{(i)}_0\right)^3},\quad& |{\bf x}-{\bf x}'|\leq r_0^{(i)}\\
		0,\quad& |{\bf x}-{\bf x}'|>r_0^{(i)}\\
	\end{array}
	\right.
\end{equation}
with the following Fourier transform:
\begin{equation} \label{eq:potentFour}
	U^{(i)}({\bf p}-{\bf p}')  = 
	3g^{(i)}\frac{j_1(|{\bf p}-{\bf p}'|/p_0^{(i)})}{|{\bf p}-{\bf p}'|/p_0^{(i)}},
	\end{equation}
where $p_0^{(i)}=\hbar/r_0^{(i)}$ and $j_{1}(x)=\sin(x)/x^{2}-\cos(x)/x$ is the spherical Bessel function. In the limit $r_{0}^{(i)}\to 0$, the potential becomes of contact type, $U^{(i)}({\bf p}-{\bf p}')=g^{(i)}=\frac{4\pi{\hbar}^{2}}{m}a^{(i)}$, where $a^{(i)}$ is the $s$-wave scattering length. The integral in Eq.~\eqref{eq:DispFun} should be computed in the sense of its principle value,
\begin{equation}
    \lim_{\varepsilon\to +0}\,\frac{1}{x+i\varepsilon
    }={\rm p.v.}\left(\frac{1}{x}\right)-i\pi\delta(x).
\end{equation}
Consequently, in general case, $Y(w,h,\kappa)$ is a complex function. Therefore, to satisfy Eq.~\eqref{eq:DispEq} we must assume $\omega$ to be a complex number, $\omega\to \omega-i\gamma$. This replacement should be done in Eq.~\eqref{eq:KinEqg}. Then it is easy to see that the resulting equation \eqref{eq:DispEq} acquires the additional term $\propto -i\gamma$. Thus, this equation is satisfied when its real and imaginary parts turn to zero.

\begin{figure}[htb]
    \centering
    \includegraphics[width=0.8\linewidth]{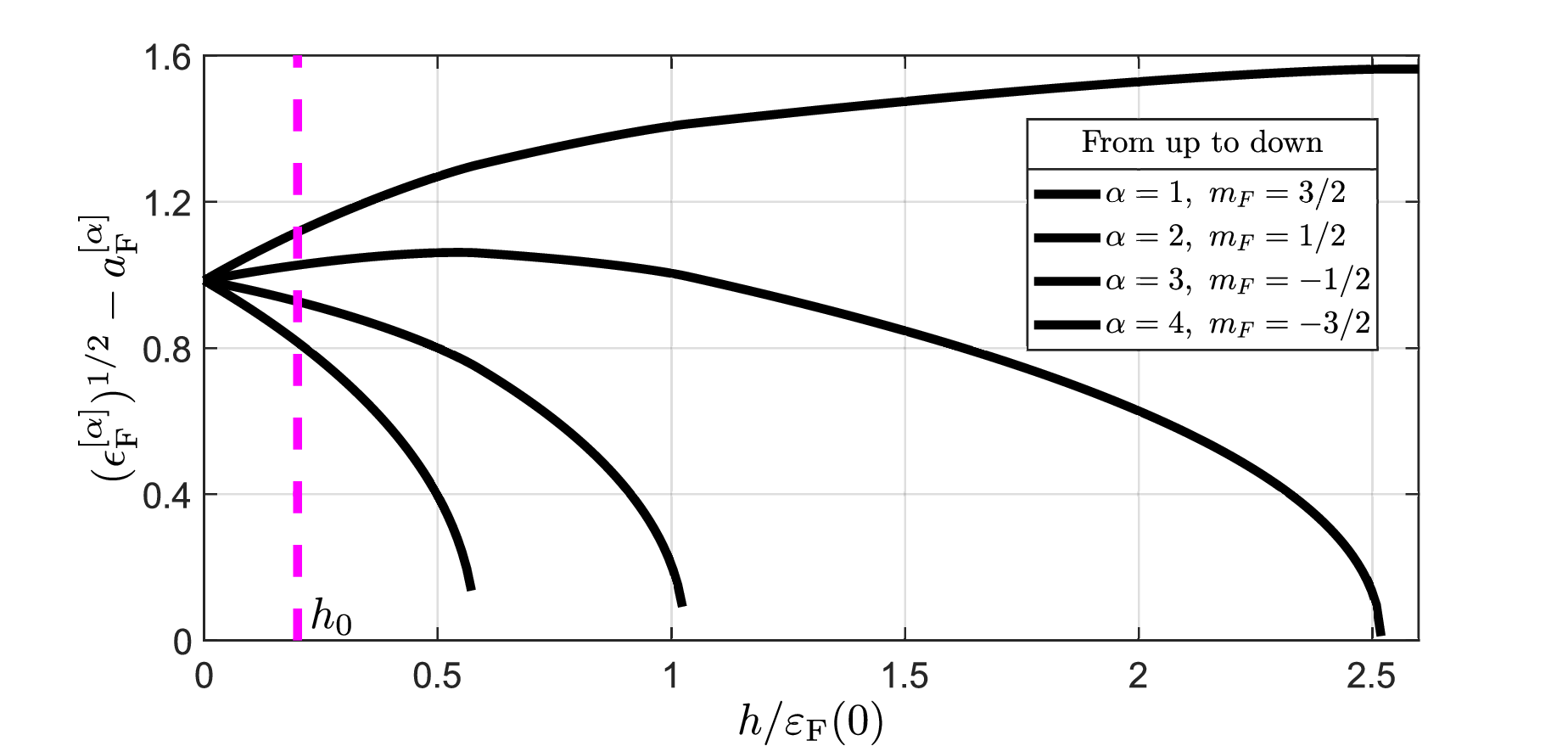}
    \includegraphics[width=0.8\linewidth]{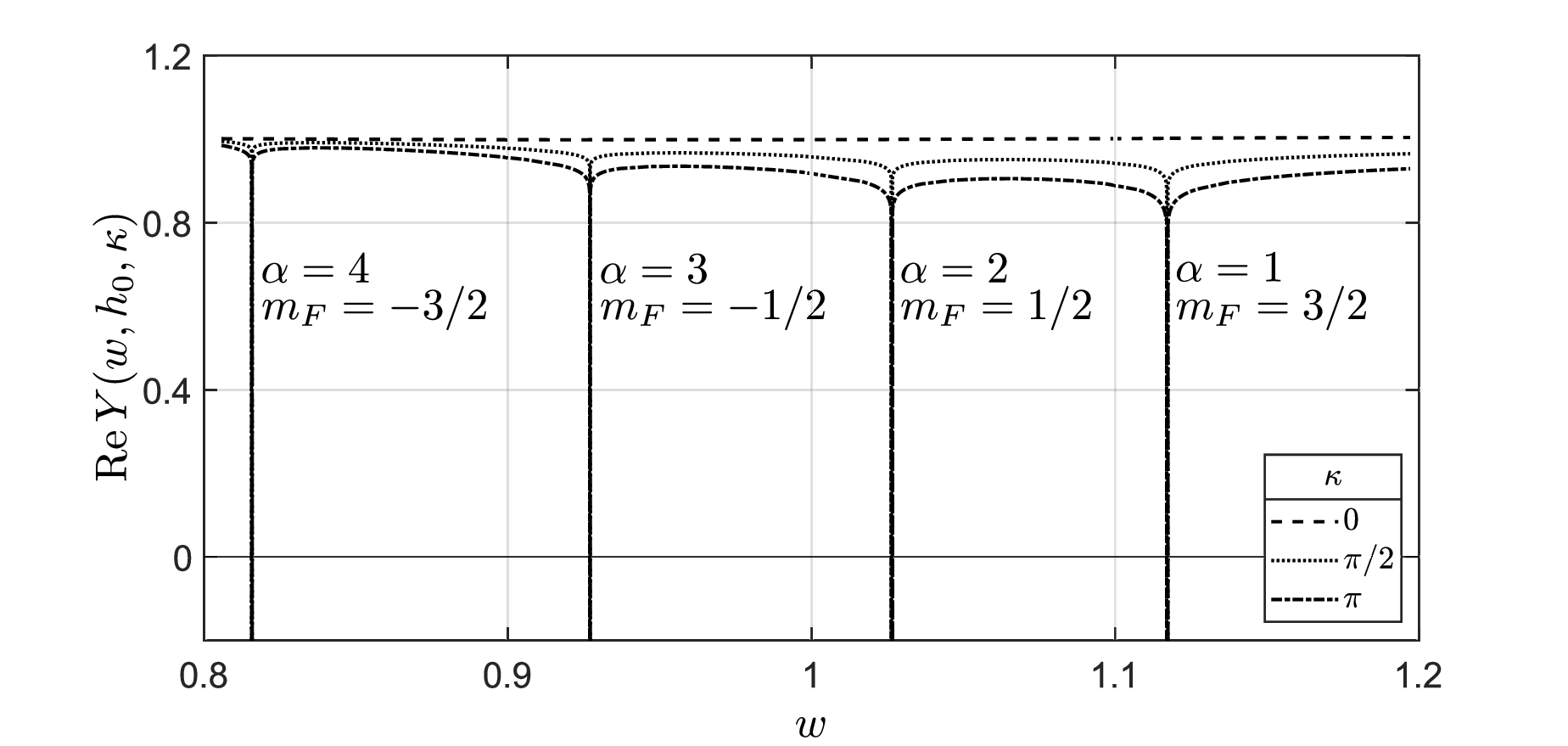}
    \caption{The dependencies of four poles of Eq.~\eqref{eq:DispFun} on the "magnetic field" $h$ (top panel). The magenta dashed line indicates the value $h_{0}$ at which we plot the function $Y(w,h_{0},\kappa)$ in the bottom panel. The solutions of Eq.~\eqref{eq:DispEq} are realized at the intersection of the function $Y(w,h_{0},\kappa)$ with a horizontal solid line. All calculations are performed for the following values of the physical parameters: $n=10^{15}\,\text{cm}^{-3}$, $a^{(i)}=100a_0$, $r_{0}^{(0)}=760a_0$, $r_{0}^{(1)}=780a_0$, $r_{0}^{(2)}=800a_0$, $r_{0}^{(3)}=820a_0$, and $h_0=0.2\varepsilon_{\rm F}(0)$, where $a_{0}\approx 53\,\text{pm}$ is the Bohr radius.}
    
    \label{fig:DispSol}
\end{figure}

First, we address the numerical analysis of the real part of the dispersion equation, see Eq.\eqref{eq:DispEq}. It determines the dimensionless speed of zero sound $w$ as a function of the "magnetic field" $h$ and angle $\kappa$. As we see, Eq.~\eqref{eq:DispFun} has four poles $w=\sqrt{\epsilon_{\rm F}^{[\alpha]}}-a_{\rm F}^{[\alpha]}$, which are realized at the lower limit of integration over $\theta$. The top panel in Fig.~\ref{fig:DispSol} shows the dependencies of all poles on the "magnetic field" $h$, which originate due to the fact that $\epsilon_{\rm F}^{[\alpha]}=\epsilon_{\rm F}^{[\alpha]}(h)$. The magenta dashed line indicates the value of $h$ at which we plot the function $Y(w)$ at different values of angle $\kappa$, as shown in the bottom panel of Fig.~\ref{fig:DispSol}.  It is clear that the solutions of Eq.~\eqref{eq:DispEq} are then realized  at the intersection of the function $Y(w)$ with a solid black line. It should be noted that these solutions are difficult to illustrate graphically since they lie very close to the poles of the integrand (for better visualization, we take large values of $r_{0}^{(i)}$ typical to Rydberg atoms). Nevertheless, in the vicinity of each pole indicated by four vertical lines there are two solutions lying on either side of them. Hence, we have eight zero sound modes. Sound modes corresponding to the solutions on the left of the poles are always slower than those on the right of them. 
Therefore, we can speak of "slow" and "fast" waves \cite{Poluektov_2014}. All solutions demonstrate a weak dependence on the angle $\kappa$. An exception is the case $\kappa=0$ in which there is no solution at all.  This can be also shown analytically from Eq.~\eqref{eq:DispFun} by expanding $U^{(i)}$ into the series.            

It is worth noting that the experimental observation of zero sound in the system under consideration allows us to judge the character of the interaction. Indeed, since the speed of zero sound lies very close to the poles (as shown in Fig.~\ref{fig:DispSol}), its measured value sets their position. In turn, the poles are determined not only by $\varepsilon_{\rm F}^{[\alpha]}$ but also by the quantity $a_{\rm F}^{[\alpha]}$ that, according to Eq.~\eqref{eq:FuncA}, characterizes the nonlocality of interaction.    

\begin{figure}
    \centering
    \includegraphics[width=0.8\linewidth]{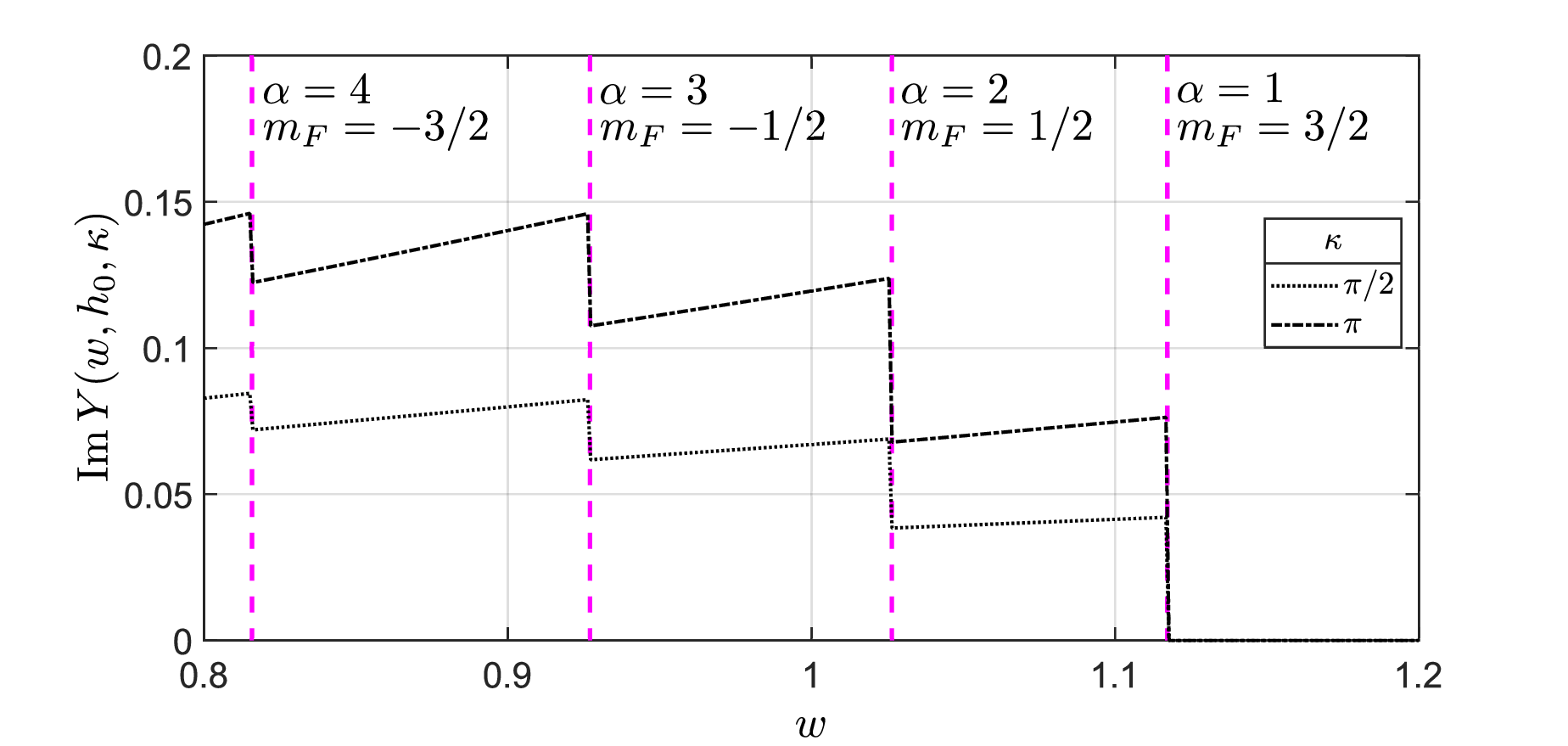}
    \caption{The dependence of the decrement $\gamma$, ${\rm Im}\,Y(w,h,\kappa)\propto \gamma$, on $w$ at fixed magnetic field $h_0=0.2\varepsilon_{\rm F}(0)$. The magenta dashed lines indicate the positions of poles. The calculations are performed for the following values of the physical parameters: $n=10^{15}\,\text{cm}^{-3}$, $a^{(i)}=100a_0$, $r_{0}^{(0)}=760a_0$, $r_{0}^{(1)}=780a_0$, $r_{0}^{(2)}=800a_0$, $r_{0}^{(3)}=820a_0$, and $h_0=0.2\varepsilon_{\rm F}(0)$, where $a_{0}\approx 53\,\text{pm}$ is the Bohr radius.}
    \label{fig:DispImag}
\end{figure}

Now we discuss the damping of zero sound due to Landau collisionless mechanism \cite{Slyusarenko_1998}. As we already mentioned, it is determined by the imaginary part of the dispersion equation \eqref{eq:DispEq}, ${\rm Im}\,Y(w,h,\kappa)\propto\gamma$. The damped oscillations with decrement $\gamma>0$ indicate the stability of the system. Figure \ref{fig:DispImag} shows the dependence of ${\rm Im}\,Y(w,h_{0},\kappa)$ on $w$ at fixed "magnetic field" $h_{0}$ and different values of $\kappa$. 
As we see, seven of eight modes (there are two solutions in the vicinity of each pole) are damped. Moreover, the "slow" modes have a larger decay rate. There is only one undamped mode ($\gamma=0$) corresponding to the spin projection $m_{F}=3/2$. 

\section{Conclusion}
\label{Summary}

In the context of quantum gases of high-spin atoms, we have obtained and justified the many-body Hamiltonian with multipole exchange interaction. All relevant multipole moments specifying this interaction are included into the Hamiltonian by means of the irreducible spherical tensor operators. 
We argue that the discussed Hamiltonian makes sense only for finite-range interatomic interaction. The zero-range (pseudo)potentials parameterized by $s$-wave scattering lengths do not 
give rise the multipole exchange interactions.


The direct observation of multipole degrees of freedom is difficult to achieve. For this reason, they are referred to as hidden parameters, which, however, could manifest themselves in the collective effects of many-body systems. 
To study this issue, we have addressed the specific system representing a degenerate gas of interacting spin-3/2 atoms in a magnetic field. We have employed the respective Hamiltonian with multipole (spin, quadrupole, octupole) exchange interactions to derive the kinetic equation in the collisionless approximation.  Next, we have applied the resulting equation to study the high-frequency oscillations known as zero sound. We have succeeded in demonstrating that zero sound emerges due to finite-range interaction, which, in turn, is rigidly related to the manifestation of multipole degrees of freedom. Moreover, by measuring the speed of zero sound, one can judge the degree of nonlocality of the interaction. 
From the analysis of the dispersion equation, we have shown that there are eight zero-sound modes: for each of the four spin projections, there are two waves -- "slow" and "fast". The "slow" waves are characterized by a larger damping factor than the "fast" ones. The only one "fast" mode corresponding to the spin projection $m_{F}=3/2$ has a zero decrement. 

We have demonstrated that within the rigorous mathematical formulation of the Hamiltonian for high-spin systems, it is necessary to take into account all processes of multipole exchange. As for the physical aspects of multipole exchange interactions, it can be proved that the quadrupole exchange interaction is related to the dipole-dipole forces \cite{Racah1959}. Therefore, for atoms with large intrinsic dipole moments, the quadrupole exchange interaction is essential \cite{PitStr2016,Bulakhov_2022}. The physical importance of octupole exchange interaction demands additional study.

\section*{Acknowledgements}
 The authors acknowledge support from the National Research Foundation of Ukraine, Grant No. 0120U104963, Ministry of Education and Science of Ukraine, Research Grant No. 0122U001575, National Academy of Sciences of Ukraine, Project No. 0121U108722, and STCU project "Magnetism for Ukraine", Grant No. 9918.

\begin{appendix}
\section{The projection operator approach}
\label{app:Projection_Op}

Let us consider a low-energy collision of two identical atoms with spin $F=3/2$. Since the total spin ${\cal F}$ is conserved (the orbital angular momentum is zero for $s$-wave scattering), their interaction can be written in the form
\begin{equation}\label{eq:PairHam}
V=\sum_{{\cal F}=0}^{3}g_{\cal F}P_{\cal F}, 
\end{equation}
where $g_{\cal F}$ is the coupling constant in the total spin ${\cal F}$ scattering channel and $P_{\cal F}$ is the projection operator 
onto a state with total spin ${\cal F}$. This operator has the following property: $P_{\cal F}P_{{\cal F}'}=P_{\cal F}\delta_{{\cal F}{\cal F}'}$. The relation 
$$
({\bf F}_{1}\cdot {\bf F}_{2})=\sum_{{\cal F}=0}^{3}\lambda_{\cal F}P_{\cal F}
$$
with
$$
\lambda_{\cal F}=\frac{1}{2}\left[{\cal F}({\cal F}+1)-2F(F+1)\right], \quad F=3/2
$$
gives
\begin{gather}
({\bf F}_{1}\cdot {\bf F}_{2})=\frac{1}{4}(-{15}P_{0}-11P_{1}-3P_{2}+9P_{3}), \nonumber \\
({\bf F}_{1}\cdot {\bf F}_{2})^{2}=\frac{1}{16}(225P_{0}+121P_{1}+9P_{2}+81P_{3}), \nonumber \\
({\bf F}_{1}\cdot {\bf F}_{2})^{3}=\frac{1}{64}(-3375P_{0}-1331P_{1}-27P_{2}+729P_{3}). \label{eq:couple_eq}
\end{gather}
To solve these equations with respect to $P_{\cal F}$, we need to add the fourth equation representing the completeness condition for the projection operator, $\sum_{\cal F}P_{\cal F}=1$. Then, the solution of Eqs.~(\ref{eq:couple_eq}) is
\begin{gather*}
P_{0}=-\frac{1}{18}({\bf F}_{1}\cdot{\bf F}_{2})^{3}-\frac{5}{72}({\bf F}_{1}\cdot{\bf F}_{2})^{2}+\frac{31}{96}({\bf F}_{1}\cdot{\bf F}_{2})+\frac{33}{128}, \\
P_{1}=\frac{1}{10}({\bf F}_{1}\cdot{\bf F}_{2})^{3}+\frac{9}{40}({\bf F}_{1}\cdot{\bf F}_{2})^{2}-\frac{117}{160}({\bf F}_{1}\cdot{\bf F}_{2})-\frac{81}{128}, \\
P_{2}=-\frac{1}{18}({\bf F}_{1}\cdot{\bf F}_{2})^{3}-\frac{17}{72}({\bf F}_{1}\cdot{\bf F}_{2})^{2}+\frac{23}{96}({\bf F}_{1}\cdot{\bf F}_{2})+\frac{165}{128}, \\
P_{3}=\frac{1}{90}({\bf F}_{1}\cdot{\bf F}_{2})^{3}+\frac{29}{360}({\bf F}_{1}\cdot{\bf F}_{2})^{2}+\frac{27}{160}({\bf F}_{1}\cdot{\bf F}_{2})+\frac{11}{128}.
\end{gather*}
Therefore, the pairwise interaction Hamiltonian, according to Eq.~(\ref{eq:PairHam}), reads
\begin{equation}\label{eq:Int_mult}
V=c_{0}+c_{1}({\bf F}_{1}\cdot {\bf F}_{2})+c_{2}({\bf F}_{1}\cdot {\bf F}_{2})^{2}+c_{3}({\bf F}_{1}\cdot {\bf F}_{2})^{3},   
\end{equation}
where
\begin{gather*}
c_{0}=\frac{1}{128}\left[33g_{0}-81g_{1}+165 g_{2}+11g_{3}\right],\\
c_{1}=\frac{1}{32}\left[\frac{31g_{0}+23g_{2}}{3}+\frac{27g_{3}-117g_{1}}{5}\right], \\
c_{2}=\frac{1}{8}\left[-\frac{5g_{0}+17g_{2}}{9}+\frac{1}{5}\left(9g_{1}+\frac{29g_{3}}{9}\right)\right],     \\
c_{3}=\frac{1}{2}\left[-\frac{g_{0}+g_{2}}{9}+\frac{1}{5}\left(g_{1}+\frac{g_{3}}{9}\right)\right].
\end{gather*}
Hence, Eq.~(\ref{eq:Int_mult}) shows that the general structure of interatomic interaction is specified by the spin-independent term, as well as by bilinear, biquadratic, and bicubic terms in spin-3/2 operators. Two latter terms are responsible for the quadrupole and octupole exchange interactions inherent in high-spin systems.  

However, if the many-body wave function is represented as the product of its orbital and spin parts, then the exchange of two identical atoms results in the fact that only even values of the total spin $\cal F$ contribute to the Hamiltonian \eqref{eq:PairHam} regardless of the quantum statistics \cite{UedaPhysRep}. Therefore, we must set $P_{1}=0$ and $P_{3}=0$. This requirement allows us to exclude, for example, the biqudratic and bicubic terms in expressions for $P_{0}$ and $P_{2}$ such that the resulting Hamiltonian contains only the terms of the zeroth and first powers of $({\bf F}_{1}\cdot {\bf F}_{2})$. The respective coupling constants can be expressed in terms of two $s$-wave scattering lengths corresponding to ${\cal F}=0$ and ${\cal F}=2$ channels. Therefore, one may conclude that the scattering-length approximation ignores the multipole exchange interactions. 

\section{Matrix equivalents of irreducible spherical tensor operator}

Although the theory of irreducible spherical tensor operators is widely discussed in the literature \cite{Racah1959,Brink_1968,Wigner_1959,Ryabov_1999}, we briefly present the main formulas used in the paper. The matrix elements of the spherical tensor operator $T_{q}^{k}$ (the upper and lower indices denote its rank and component, respectively) are given by
\label{app:STO}
\begin{equation}
    \left<j\,m|T^k_q|j'\,m'\right>
    =
    (-1)^{j-m}
    \left(
    \begin{array}{w{c}{1.6em}w{c}{1.6em}w{c}{1.6em}}
        j & k & j'\\
        -m & q & m'
    \end{array}
    \right)
    \left<j||T^k||j'\right>
    ,
\end{equation}
where the 2 × 3 array represents the 3-j Wigner symbol, $\left<j||T^k||j'\right>=\sqrt{2k+1}$ is the Racah's normalization and $|j\,m\rangle$ represents the eigenvector of the  angular momentum operator of a particle.
This expression can be employed to generate the following matrix equivalents of the spherical tensor operators for $j=3/2$:
\begin{align*}
&&T^0_0
=
\frac12
\left(
\begin{array}{w{c}{3ex}w{c}{3ex}w{c}{3ex}w{c}{3ex}}
 1 & 0 & 0 & 0 \\
 0 & 1 & 0 & 0 \\
 0 & 0 & 1 & 0 \\
 0 & 0 & 0 & 1 \\
\end{array}
\right)
,&\hfill
&T^1_{1}
=
-\frac{1}{\sqrt{10}}
\left(
\begin{array}{w{c}{3ex}w{c}{3ex}w{c}{3ex}w{c}{3ex}}
 0 & \sqrt{3} & 0 & 0 \\
 0 & 0 & 2 & 0 \\
 0 & 0 & 0 & \sqrt{3} \\
 0 & 0 & 0 & 0 \\
\end{array}
\right)
,
\\[1ex]
&&T^1_{-1}
=
\frac{1}{\sqrt{10}}
\left(
\begin{array}{w{c}{3ex}w{c}{3ex}w{c}{3ex}w{c}{3ex}}
 0 & 0 & 0 & 0 \\
 \sqrt{3} & 0 & 0 & 0 \\
 0 & 2 & 0 & 0 \\
 0 & 0 & \sqrt{3} & 0 \\
\end{array}
\right)
,&\hfill
&T^1_{0}
=
\frac{1}{\sqrt{5}}
\left(
\begin{array}{w{c}{3ex}w{c}{3ex}w{c}{3ex}w{c}{3ex}}
 \dfrac{3}{2} & 0 & 0 & 0 \\
 0 & \dfrac{1}{2} & 0 & 0 \\
 0 & 0 & -\dfrac{1}{2} & 0 \\
 0 & 0 & 0 & -\dfrac{3}{2} \\
\end{array}
\right)
,
\\[1ex]
&
&T^2_{-2}
=
\sqrt{\frac{1}{2}}
\left(
\begin{array}{w{c}{3ex}w{c}{3ex}w{c}{3ex}w{c}{3ex}}
 0 & 0 & 0 & 0 \\
 0 & 0 & 0 & 0 \\
 1 & 0 & 0 & 0 \\
 0 & 1 & 0 & 0 \\
\end{array}
\right)
,
&\hfill
&T^2_{-1}
=
\sqrt{\frac{1}{2}}
\left(
\begin{array}{w{c}{3ex}w{c}{3ex}w{c}{3ex}w{c}{3ex}}
 0 & 0 & 0 & 0 \\
 1 & 0 & 0 & 0 \\
 0 & 0 & 0 & 0 \\
 0 & 0 & -1 & 0 \\
\end{array}
\right)
,
\\[1ex]
&
&T^3_{-3}
=
\left(
\begin{array}{w{c}{3ex}w{c}{3ex}w{c}{3ex}w{c}{3ex}}
 0 & 0 & 0 & 0 \\
 0 & 0 & 0 & 0 \\
 0 & 0 & 0 & 0 \\
 1 & 0 & 0 & 0 \\
\end{array}
\right)
,
&\hfill
&T^3_{-2}
=
\sqrt{\frac{1}{2}}
\left(
\begin{array}{w{c}{3ex}w{c}{3ex}w{c}{3ex}w{c}{3ex}}
 0 & 0 & 0 & 0 \\
 0 & 0 & 0 & 0 \\
 1 & 0 & 0 & 0 \\
 0 & -1 & 0 & 0 \\
\end{array}
\right)
,
\\[1ex]
&&T^2_{2}
=
\sqrt{\frac{1}{2}}
\left(
\begin{array}{w{c}{3ex}w{c}{3ex}w{c}{3ex}w{c}{3ex}}
 0 & 0 & 1 & 0 \\
 0 & 0 & 0 & 1 \\
 0 & 0 & 0 & 0 \\
 0 & 0 & 0 & 0 \\
\end{array}
\right)
,&\hfill
&T^3_{3}
=
\left(
\begin{array}{w{c}{3ex}w{c}{3ex}w{c}{3ex}w{c}{3ex}}
 0 & 0 & 0 & -1 \\
 0 & 0 & 0 & 0 \\
 0 & 0 & 0 & 0 \\
 0 & 0 & 0 & 0 \\
\end{array}
\right)
,\\[1ex]
&
&T^2_{1}
=
\sqrt{\frac{1}{2}}
\left(
\begin{array}{w{c}{3ex}w{c}{3ex}w{c}{3ex}w{c}{3ex}}
 0 & -1 & 0 & 0 \\
 0 & 0 & 0 & 0 \\
 0 & 0 & 0 & 1 \\
 0 & 0 & 0 & 0 \\
\end{array}
\right)
,
&\hfill
&T^3_{2}
=
\sqrt{\frac{1}{2}}
\left(
\begin{array}{w{c}{3ex}w{c}{3ex}w{c}{3ex}w{c}{3ex}}
 0 & 0 & 1 & 0 \\
 0 & 0 & 0 & -1 \\
 0 & 0 & 0 & 0 \\
 0 & 0 & 0 & 0 \\
\end{array}
\right)
,
\\[1ex]
&
&T^2_{0}
=
\frac{1}{2}
\left(
\begin{array}{w{c}{3ex}w{c}{3ex}w{c}{3ex}w{c}{3ex}}
 1 & 0 & 0 & 0 \\
 0 & -1 & 0 & 0 \\
 0 & 0 & -1 & 0 \\
 0 & 0 & 0 & 1 \\
\end{array}
\right)
,
&\hfill
&T^3_{1}
=
\frac{1}{\sqrt{5}}
\left(
\begin{array}{w{c}{3ex}w{c}{3ex}w{c}{3ex}w{c}{3ex}}
 0 & -1 & 0 & 0 \\
 0 & 0 & \sqrt{3} & 0 \\
 0 & 0 & 0 & -1 \\
 0 & 0 & 0 & 0 \\
\end{array}
\right)
,
\\[1ex]
&
&T^3_{-1}
=
\frac{1}{\sqrt{5}}
\left(
\begin{array}{w{c}{3ex}w{c}{3ex}w{c}{3ex}w{c}{3ex}}
 0 & 0 & 0 & 0 \\
 1 & 0 & 0 & 0 \\
 0 & -\sqrt{3} & 0 & 0 \\
 0 & 0 & 1 & 0 \\
\end{array}
\right)
,
&\hfill
&T^3_{0}
=
\frac{1}{\sqrt{5}}
\left(
\begin{array}{w{c}{3ex}w{c}{3ex}w{c}{3ex}w{c}{3ex}}
 \dfrac{1}{2} & 0 & 0 & 0 \\
 0 & -\dfrac{3}{2} & 0 & 0 \\
 0 & 0 & \dfrac{3}{2} & 0 \\
 0 & 0 & 0 & -\dfrac{1}{2} \\
\end{array}
\right)
.
\end{align*}
Let us also enumerate some valuable properties of the above matrices:
\begin{itemize}
    \item Normalization condition
\begin{equation}\label{eq:Norm}
\Tr T_{m}^{i}T^{i'}_{m'}=(-1)^{m}\delta_{i,i'}\delta_{m,-m'}
.
\end{equation}
\item Fierz identity
\begin{equation}
    \sum_{i=0}^3
    \sum_{m=-i}^{i}
    (-1)^m(T_{m}^{i})_{\alpha\beta}(T^{i}_{-m})_{\gamma\sigma}
    =
    \delta_{\alpha\sigma}
    \delta_{\gamma\beta}.
    \end{equation}

\item Fierz-like identity
\begin{equation}
    \sum_{m=-i}^i
    (-1)^m
    (T_{m}^{i})_{\alpha\beta}
    (T_{k}^{j})_{\beta\gamma}
    (T^{i}_{-m})_{\gamma\sigma}
    =
    A^{ij}
    (T_{k}^{j})_{\alpha\sigma},
\end{equation}
where
\begin{equation*}
    A^{ij}
    =
    \frac14
    \left(
        \begin{array}{cccc}
            1 & 1 & 1 & 1 \\[2ex]
            3 & \dfrac{11}{5} & \dfrac{3}{5} & -\dfrac{9}{5} \\[2ex]
            5 & 1 & -3 & 1 \\[2ex]
            7 & -\dfrac{21}{5} & \dfrac{7}{5} & -\dfrac{1}{5} \\[2ex]
        \end{array}
    \right)
    .
\end{equation*}
\item Selection rule
\begin{equation}
\begin{gathered}
        (T_{m_1}^{i_1})_{\alpha\beta}
    (T_{m_2}^{i_2})_{\beta\gamma}
    =
    \sum_{j=|i_1-i_2|}^{i_1+i_2}
    B(i_1,i_2,m_1,m_2;j,k)
    (T^j_k)_{\alpha\gamma}
    ,\\
    k=m_1+m_2
    ,\ 
    |k|\leq j\leq 3.
\end{gathered}
\end{equation}
\end{itemize}
\end{appendix}

\section*{References}
\bibliographystyle{iopart-num}
\bibliography{main}

\end{document}